\documentclass[prb,unsortedaddress,twocolumn,showpacs,showkeys]{revtex4}
\usepackage{graphicx}
\usepackage{amsmath}
\usepackage{amsfonts}
\usepackage{amssymb}

\newcommand{\res}{\textrm{res}}
\newcommand{\dx}{dx}
\newcommand{\FPS}{\textrm{FPS}}
\newcommand{\vecx}{\mathbf{x}}
\newcommand{\RR}{\mathbb{R}}

\begin{document}

\title{Rigorous derivation of coherent resonant tunneling time and velocity in finite periodic systems}
\author{C. Pacher, W. Boxleitner and E. Gornik}
\affiliation{Institut f\"ur Festk\"orperelektronik, 
Technische Universit\"at Wien, A-1040 Wien,
Austria} \affiliation{ARC Seibersdorf Research GmbH, Donau-City
Stra\ss e 1/4, A-1220 Wien, Austria}
\date{\today}
\begin{abstract}
The velocity $v_{\res}$ of resonant tunneling electrons in
finite periodic structures is analytically calculated in two ways.
The first method is based on the fact that a transmission of unity
leads to a coincidence of all still competing tunneling time
definitions. Thus, having an indisputable resonant tunneling time
$\tau_{\res},$ we apply the natural definition
$v_{\res}=L/\tau_{\res}$ to calculate the velocity. For the second
method we combine Bloch's theorem with the transfer matrix
approach to decompose the wave function into two Bloch waves. Then
the expectation value of the velocity is calculated. Both
different approaches lead to the same result, showing their
physical equivalence. The obtained resonant tunneling velocity
$v_{\res}$ is smaller or equal to the group velocity times the
magnitude of the complex transmission amplitude of the unit cell.
Only at energies where the \emph{unit cell} of the periodic
structure has a transmission of unity $v_{\res}$ equals the group
velocity. Numerical calculations for a GaAs/AlGaAs superlattice
are performed. For typical parameters the resonant velocity is below
one third of the group velocity.
\end{abstract}
 \maketitle

\section{Introduction}

There has been an ongoing debate about the time an electron spends when it
passes through a classically forbidden region (e.g. a rectangular barrier) for
many decades. Despite a number of review articles\cite{Hauge89,Landauer94} and
many papers up to now there still exist different definitions of tunneling
time. Most of the studies have been performed in one of the following
frameworks: (i) wave packet analysis
\cite{Wigner55,Hauge87,Leavens89,Hauge89,Marinov97}, (ii) dynamic paths
including Feynman paths \cite{Sokolovski87,Fertig90,Sokolovski93}, (iii)
physical
clocks\cite{Buettiker82,Buettiker83,Falck88,Leavens89b,Martin93,Gasparian93,Gasparian95,Li02}%
, (iv) flux-flux correlation functions\cite{Pollak84}, (v) theory of weak
measurements\cite{Steinberg95}, and (vi) combinations of the
former\cite{Sokolovski90}.

The real phase or delay time
\begin{equation}
\tau_{\textrm{phase}}(E)=\frac{\partial\arg t(E)}{\partial\omega}=\hbar\frac
{\partial\arg t(E)}{\partial E}, \label{defphasetime}%
\end{equation}
where $t(E)$ is the complex transmission coefficient as a function of energy,
and $\omega$ is the angular frequency, arises from a stationary-phase argument
for the transmitted wave packet. Many of the approaches (ii) to (vi) result at
first in complex transmission tunneling times, given by one of the following
expressions:%
\begin{align}
\tau_{T}^{E}(E)  &  =-i\hbar\frac{\partial\ln t(E)}{\partial E}=\tau
_{\textrm{phase}}(E)-i\hbar\frac{\partial\ln|t(E)|}{\partial E},\label{tTE}\\
\tau_{T}^{V}(E)  &  =i\hbar\frac{\partial\ln t(E)}{\partial V},\label{tTV}\\
\tau_{T}^{\delta V}(E)  &  =i\hbar\frac{\delta\ln t(E)}{\delta V(x)}=\tau
_{T}^{E}(E)-i\hbar\frac{r(E)+r^{\prime}(E)}{4E}, \label{tTdV}%
\end{align}
where $\delta/\delta V(x)$ denotes the functional derivative with respect to
the potential $V(x),\;$and $r(E)$ and $r^{\prime}(E)\;$are the reflection
amplitudes for particles coming from the left and right side, respectively.
The corresponding reflection times $\tau_{R}^{X}(E)$ are given by the substitution of $t(E)$
by $r(E)$ in the left equalities in Eqs.~(\ref{tTE}) to (\ref{tTdV}). In the
framework of physical
clocks\cite{Buettiker82,Buettiker83,Falck88,Leavens89b,Martin93,Gasparian93,Gasparian95,Li02}
the resulting times have been the real or imaginary part or the absolute value
of one of the times in (\ref{tTE}) and (\ref{tTV}).

On the other hand, it was shown that the simultaneous process of (a) the determination whether a particle is transmitted and (b) if so, how long it
took to traverse the barrier, corresponds to two non-commuting
observables\cite{Dumont93}. There was also some direct evidence that the
imaginary part of the tunneling time results from the backaction on the
particle due to the measurement process\cite{Steinberg95}.

In contrast, the time an electron spends under the barrier, either finally
reflected by or transmitted through the barrier, is consistently given by the
dwell time\cite{Smith60}, also called sojourn time, which is defined as the
ratio of the number of particles within the barrier (extending from $a$ to
$b$) to the incident flux\cite{Buettiker83}:%
\begin{equation}
\tau_{D}(E)=\frac{1}{v_{in}}\int_{a}^{b}|\Psi|^{2}dx. \label{dwelltimedef}%
\end{equation}
A Hermitian sojourn time operator exists\cite{Jaworski89}, which shows that
this time is measurable.

Again there has been no general agreement whether a relation of the following
form must hold:%
\begin{equation}
\tau_{D}(E)=|t(E)|^{2}\tau_{T}(E)+|r(E)|^{2}\tau_{R}(E). \label{dwellassum}%
\end{equation}
Based on the argument that reflection and transmission are mutually exclusive
events, that exhaust all possibilities in the sense of Feynman, this relation
served as a point of focus in an early review\cite{Hauge89}. The complex
tunneling times $\tau_{T,R}^{\delta V}$ fulfill Eq.~(\ref{dwellassum}%
).\cite{Sokolovski87} Nevertheless, the arguments for Eq.~(\ref{dwellassum})
have also been critisized arguing that the used approach goes beyond Feynman's
original interpretation\cite{Landauer94}.

\subsection*{Tunneling times in case of transmission equals unity}

Given that $|t|=1,$ for a certain energy, it was shown that phase time and
dwell time are identical, not only for the single barrier (at energies higher
than the potential energy of the barrier)\cite{Buettiker83}, but also for
arbitrary structures\cite{Leavens89,Hauge89}. The tunneling times $\tau
_{T}^{E},\tau_{T}^{V}$ and $\tau_{T}^{\delta V},$ Eqs.~(\ref{tTE}) to
(\ref{tTdV}), also simplify to the phase time $\tau_{\textrm{phase}},$ Eq.
(\ref{defphasetime}), for $|t|=1$ in any arbitrary structure. Thus we have%
\begin{align}
|t(E^{\prime})|=1\Rightarrow & \nonumber\\
\tau_{\textrm{phase}}(E^{\prime})=  &  \tau_{D}(E^{\prime})=\tau_{T}^{E}(E^{\prime
})=\tau_{T}^{V}(E^{\prime})=\tau_{T}^{\delta V}(E^{\prime}).
\label{equalityofalltaus}%
\end{align}
In accordance with the real character of the phase time the aforementioned
problem of non-commuting observables vanishes in the case $|t(E)|=1,$ since
all particles tunnel finally through the structure; there is neither
reflection nor interference between reflected and transmitted particles.

Stimulated by comments in Ref. \onlinecite{Landauer94} and recent
results from the theory of finite periodic systems
\cite{Pereyra02,Pacher03}, here we study the tunneling time of electrons
that tunnel resonantly through finite periodic systems. In contrast to
tunneling through single barriers, periodic systems have a transmission
probability of unity below the barrier potential at the individual
transmission resonances, which form allowed energy bands. Due to
Eq.~(\ref{equalityofalltaus}) we can choose any time definition, Eqs.
(\ref{defphasetime}) to (\ref{dwelltimedef}), to calculate the tunneling
time at resonance. We will use the phase delay time,
Eq.~(\ref{defphasetime}), and make use of the results obtained in Refs.
\onlinecite{Pacher03} and \onlinecite{Pereyra00} recently.

At first view, a similiar approach to the phase time is the concept of the
group velocity%
\begin{equation}
v_{g}=\frac{\partial\omega}{\partial k},
\end{equation}
which is the velocity of the envelope of a propagating wave packet in a
medium. Here $\omega$ is the angular frequency and $k$ is the wave number. The
function $\omega(k)$ is normally referred to as dispersion. The solutions of
the Schr\"{o}dinger equation for a periodic potential also yield a (band)
dispersion relation between the Bloch wave number $q$ and the angular
frequency $\omega$. Using $E=\hbar\omega$, the group velocity then reads:%
\begin{equation}
v_{g}=\left(  \hbar\frac{\partial q}{\partial E}\right)  ^{-1}.
\label{groupvel}%
\end{equation}
In Ref. \onlinecite{Landauer94} the following relation between the 
tunneling time and the group velocity was given (for the tight binding limit), 
neglecting terms due to the matching of the wave functions at the
ends of the system:
\begin{equation}
v_{g}\cong L/|\tau_{T}^{E}|. \label{vgcaL/t}%
\end{equation}
Here $L$ is the length of the periodic system. Recent numerical
calculations for finite periodic systems embedded in regions of
constant potential showed that the equality in (\ref{vgcaL/t}),
\emph{does not hold} in general\cite{Merc03}. At first glance it
might seem paradox that the group velocity and the phase delay
time, which are both based on a wave packet approach, lead to
conflicting results.

The paper is organized as follows. In section \ref{sec:FPS} we introduce the
transfer matrix approach, which is used in \ref{sec:veltime} to calculate the
resonant tunneling time and the corresponding velocity. In section
\ref{sec:expectvel} Bloch's theorem is used together with the transfer matrix
to decompose the wave function inside the finite periodic system (FPS) in two
Bloch waves. It is shown that the velocity operator must have real expectation
values at transmission resonances. Then the velocity expectation value is
explicitly calculated. In section \ref{sec:identity} we note that both velocity
approaches lead to the same result and derive an upper bound for the velocity.
The calculations show that Eq.~(\ref{vgcaL/t}) taken from Ref.
\onlinecite{Landauer94} gives a wrong estimation when the unit cell
transmission amplitude is small.

Further the special case when resonant tunneling velocity and group
velocity are identical is discussed.

Our results are not restricted to a tight binding model, but are
exact as long as only coherent transport is considered.

Finally, we apply the analytical results to a semiconductor superlattice
and illustrate them with a compilation of graphs in section
\ref{sec:GaAs-SL}.

\section{Finite periodic systems} \label{sec:FPS} 
\begin{figure}
\includegraphics[width=8.5cm]{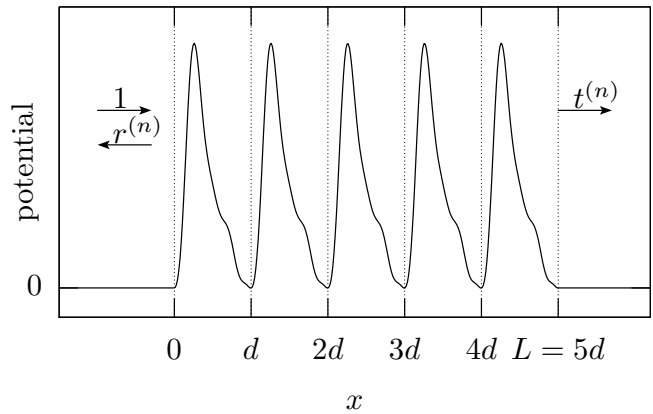}
\caption{\label{Fig:potential}Periodic potential (drawn for five periods)
embedded between two infinite half-spaces. Arrows denote the wave function (plane waves).} 
\end{figure}

Our one-dimensional
model system consists of an $n$-fold periodic structure, extending from
$x=0$ to \mbox{$x=L=nd$} ($d$ is the length of the unit cell), embedded
between two semi-infinite half spaces with zero potential (Fig.
\ref{Fig:potential}).

Assuming a plane wave, $\exp(ikx),$ travelling from the left
towards the finite periodic
system (FPS), the wave function is given by%
\begin{equation}
\Psi(x)=\left\{
\begin{array}
[c]{cc}%
\exp(ikx)+r^{(n)}\exp(-ikx)\quad &  x\leq0,\\
\Psi_{\text{FPS}}^{(n)}(x) & 0\leq x\leq L=nd,\\
t^{(n)}\exp[ik(x-L)] & x\geq L,
\end{array}
\right.  \label{Psi}%
\end{equation}
where $k=\sqrt{2mE}/\hbar$ is the electron wave vector in the semi-infinite half
spaces, and $r^{(n)}$ and $t^{(n)}$ are the complex reflection and transmission
coefficients of the $n$-fold periodic structure, respectively.

We start by briefly reviewing some important properties of one-dimensional
finite periodic systems \cite{Pereyra02,Pacher03}. The wavefunctions at the
left and right interface of a certain region, $\Psi_{L}$ and $\Psi_{R}$,
respectively, are related by the transfer matrix $M$ through\cite{LMRcomment}
$\left(
\begin{array}
[c]{c}%
A_{L}^{+}\\
A_{L}^{-}%
\end{array}
\right)  =M\left(
\begin{array}
[c]{c}%
A_{R}^{+}\\
A_{R}^{-}%
\end{array}
\right)  ,$ where $\Psi_{L,R}=(1\quad 1) \left(
\begin{array}
[c]{c}%
A_{L,R}^{+}\\
A_{L,R}^{-}%
\end{array}
\right)=A_{L,R}^{+}+A_{L,R}^{-}$. Neglecting spin, the time
reversal invariance and the conservation of the probability
density current lead to the structure\cite{Erdoes82} of the
transfer matrix $M$
\begin{equation}
M=\left(
\begin{array}
[c]{cc}%
a & b\\
b^{\ast} & a^{\ast}%
\end{array}
\right), \label{transfer matrix}%
\end{equation}
where additionally $\det M=1$ holds ($x^{\ast}$ denotes the complex conjugate
of $x$).
In terms of the transmission and reflection coefficients $t$ and $r$, the
transfer matrix can be written as\cite{Sprung93}
\begin{equation}
M=\left(
\begin{array}
[c]{cc}%
1/t & r^{\ast}/t^{\ast}\\
r/t & 1/t^{\ast}%
\end{array}
\right)  . \label{transfermatrixrt}%
\end{equation}
Since by construction the transfer matrix of a sequence of layers is the
product of the transfer matrices of each layer, the transfer matrix of a
potential consisting of $n$ periods is the $n$-th power of the transfer matrix
of one period:%

\begin{equation}
M^{n}=\left(
\begin{array}
[c]{cc}%
a^{(n)} & b^{(n)}\\
b^{(n)\ast} & a^{(n)\ast}%
\end{array}
\right)  . \label{defM^ {n}}%
\end{equation}
For $n\geq2,$ $a^{(n)}$ and $b^{(n)}$ can be expanded to\cite{Born80}
\begin{align}
a^{(n)}  &  =aU_{n-1}(\operatorname{Re}\{a\})-U_{n-2}(\operatorname{Re}%
\{a\}),\label{alpha^{n}}\\
b^{(n)}  &  =bU_{n-1}(\operatorname{Re}\{a\}), \label{beta^{n}}%
\end{align}
where $U_{n}(x)$ denote the Chebyshev Polynomials of the second kind. The
transmission $T^{(n)}$ of any (field-free) $n$-fold periodic structure is
given by\cite{Vezetti86,Yamamoto89,Sprung93,Pereyra02,Pacher03}%

\begin{equation}
T^{(n)}=\left|  a^{(n)}\right|  ^{-2}=\left[  1+|b|^{2}U_{n-1}^{2}%
(\operatorname{Re}\{a\})\right]  ^{-1}. \label{T^{(n)}}%
\end{equation}
Resonances with $T^{(n)}=1$ occur if and only if $b^{(n)}=bU_{n-1}%
(\operatorname{Re}\{a\})=0$. This leads to the condition of the transmission
resonances\cite{Pacher01,Pereyra02}
\begin{equation}
T^{(n)}=1\Leftarrow\operatorname{Re}\{a\}=\cos(j\pi/n),\quad j=1,\ldots,n-1.
\label{resonance finite SL}%
\end{equation}
The corresponding Bloch wave vectors $q_j^{\res}$ are given by the condition
$\operatorname{Re}\{a\}=\cos qd$:%
\begin{equation}
q_{j}^{\res}=\pm \frac{j\pi}{nd},\quad j=1,\dots,n-1. \label{qatresonance}%
\end{equation}
Inserting Eq.~(\ref{resonance finite SL}) into (\ref{alpha^{n}}) we get at
resonance%
\begin{equation}
a^{(n)}(q_{j}^{\res})=t^{(n)}(q_{j}^{\res})=(-1)^{j}. \label{at=(-1)j}%
\end{equation}

\section{Velocity from tunneling time}
\label{sec:veltime} Recently the phase time for a system with $n$
periods, $\tau_{\textrm{phase}}^{(n)},$ has been calculated with the help
of Eqs.~(\ref{defphasetime}), (\ref{transfermatrixrt}), and
(\ref{alpha^{n}})\cite{Pereyra00,Pacher03}:
\begin{align}
&  \tau_{\textrm{phase}}^{(n)}=\hbar T^{(n)}\left[  \left(  n-\frac{\operatorname{Re}%
\{a\}}{2}U_{2n-1}(\operatorname{Re}\{a\})\right)  \right. \label{tau^{(n)}}\\
&  \left.  \times\frac{\operatorname{Im}\{a\}}{1-\operatorname{Re}^{2}%
\{a\}}\frac{\partial\operatorname{Re}\{a\}}{\partial E}-\frac{1}{2}%
U_{2n-1}(\operatorname{Re}\{a\})\frac{\partial\operatorname{Im}\{a\}}{\partial
E}\right]  ,\nonumber
\end{align}
where $T^{(n)}$ is the transmission probability of the periodic structure
given by Eq.~(\ref{T^{(n)}}). Here we are only interested in the phase time at
resonance energies of the transmission, where the Bloch wave vector
$q_j^{\res}=j\pi/nd$ and $T^{(n)}=1.$ For energies where $T^{(n)}=1$ the phase
time is equal to the tunneling time as we discussed in the introduction and
Eq.~(\ref{tau^{(n)}})\ reduces to the in-resonance phase
time\cite{Pacher03} or resonant tunneling time:%
\begin{equation}
\tau_{\res}^{(n)}=\hbar n\frac{\operatorname{Im}\{a\}}{1-\operatorname{Re}%
^{2}\{a\}}\frac{\partial\operatorname{Re}\{a\}}{\partial E}.
\label{tauresn}
\end{equation}
Clearly, $\tau_{\res}^{(n)}$ is proportional to the number of periods $n$.
Using the natural definition
\begin{equation}
v_{\res}=L/\tau_{\res}^{(n)}, \label{vres=Lovertau}%
\end{equation}
we get the following resonant tunneling velocity
\begin{equation}
v_{\res}=\hbar^{-1}d\frac{1-\operatorname{Re}^{2}\{a\}}{-\operatorname{Im}%
\{a\}}\left(  -\frac{\partial\operatorname{Re}\{a\}}{\partial E}\right)
^{-1}, \label{resonancevelocity}%
\end{equation}
which does not depend on the number of periods $n.$ For the sake of
completeness we give the result in terms of the transmission amplitude $t=1/a$
of the unit cell:%
\begin{equation}
v_{\res}=\hbar^{-1}d\frac{|t|^{4}-\operatorname{Re}^{2}\{t\}}{\operatorname{Im}%
\{t\}}\left(  \mathbf{-}\frac{\partial\operatorname{Re}\{t\}}{\partial
E}\right)  ^{-1}. \label{vres_t}%
\end{equation}
Now it is interesting to compare this tunneling velocity $v_{\res}$ to the
group velocity $v_{g}$. From the dispersion relation $\operatorname{Re}%
\{a\}=\cos(qd)$ we obtain
\begin{equation}
v_{g}=\left(  \hbar\frac{\partial q}{\partial E}\right)  ^{-1}=\hbar
^{-1}d\sqrt{1-\operatorname{Re}^{2}\{a\}}\left(  -\frac{\partial
\operatorname{Re}\{a\}}{\partial E}\right)  ^{-1}. \label{vgroup}%
\end{equation}
Using Eq.~(\ref{resonancevelocity}), the resonant tunneling velocity and the
group velocity are related by%
\begin{equation}
v_{\res}=\frac{\sqrt{1-\operatorname{Re}^{2}\{a\}}}{-\operatorname{Im}%
\{a\}}v_{g}=\frac{\sqrt{|t|^4-\operatorname{Re}^{2}\{t\}}}{\operatorname{Im}%
\{t\}}v_{g}. \label{ratiovresvg}%
\end{equation}

Note that the equations for the resonant tunneling time and the resonant
velocity are only meaningful if they are evaluated at
energies where the FPS has a transmission probability of unity. 

Nevertheless, by increasing the number of periods towards infinity the discrete
curves tend to continuous ones, just like the dispersion relation does.

\section{Expectation value of velocity operator}
\label{sec:expectvel} In the following we calculate the
expectation value of the velocity operator applied to the exact
wave function inside the FPS.

\subsection{Decomposition of $\Psi$ into two Bloch waves}

In 1929 Bloch calculated the eigenfunctions of the Schrodinger
equation for crystal lattices\cite{Bloch29}. He modelled the
lattice by a periodic potential which spans the whole space. Using
group theory together with periodic (Born von K\'{a}rm\'{a}n)
boundary conditions he proved that the base solutions of the
Schr\"odinger equation for a lattice potential, i.e.
$V(\vecx)=V(\vecx+\mathbf{R}_i)$, where $\mathbf{R}_i$ is any
lattice vector, are of the form:
\begin{equation}
\Psi_{\mathbf{q}}^{B}(\mathbf{x})= u_{\mathbf{q}}(\mathbf{x})\exp
(i\mathbf{q}\cdot\mathbf{x}). \label{Blochwave}
\end{equation}  Here $\mathbf{q}$ is
in modern terms a reciprocal lattice vector and
\begin{equation}
u_{\mathbf{q}}(\mathbf{x}) =
u_{\mathbf{q}}(\mathbf{x}+\mathbf{R}_i)
\end{equation} is a lattice periodic function.
This fact is well known as Bloch's theorem. Since Bloch's theorem
was originally derived for an infinite domain, this infinite domain is
sometimes considered as necessary.

Therefore, we will start by briefly showing that Bloch's theorem
gives also the exact wave functions in finite systems.

Mathematically spoken, for the case of the infinite periodic
potential $V$:$\RR^3\rightarrow\RR$ considered by Bloch, the domain
of the Schr\"odinger differential equation and its solutions
$\Psi_{\mathbf{q}}^{B}$ is $D=\mathbb{R}^3$. If we reduce the
domain of the periodic potential (and thus of the Schr\"odinger
equation) to any finite subdomain $\tilde{D}\subset \mathbb{R}^3$
and choose an arbitrary potential in the domain
$\mathbb{R}^3\backslash \tilde{D}$ the base solutions of the
differential equation inside $\tilde{D}$ are not changed. The
physical consequence is that despite the loss of the global
translation invariance of the Hamiltonian in a finite system, the
wave functions are still \textit{exactly} given by superpositions
of Bloch waves, Eq.~(\ref{Blochwave}).

If we neglect spin the time inversion symmetry of the Hamiltonian
leads to Kramers degeneracy, i.e. $E(\mathbf{q})=E(-\mathbf{q})$.

Therefore, the two linear independent solutions of Schr\"odinger's
equation for the one-dimensional periodic potential in a  finite
domain are given by $\Psi_{q}^{B}$ and
$(\Psi_{q}^{B})^{\ast}(x)=\Psi_{-q}^{B}(x)$. Having clarified this
we use the following ansatz for the wave function in any
one-dimensional $n$-fold periodic system inside an allowed band:
\begin{align}
\Psi_{\text{FPS}}^{(n)}(x)= &  \alpha_{q}^{(n)}u_{q}(x)\exp(iqx)\nonumber\\
&  +\alpha_{-q}^{(n)}u_{q}^{\ast}(x)\exp(-iqx),\quad0\leq x\leq
nd\label{PsiSLnbloch}\\
u_{q}(x+d) &  =u_{q}(x),\quad u_{q}^{\ast}(x)=u_{-q}(x).\nonumber
\end{align}
The dimensionless coefficients $\alpha_{q}^{(n)}$ and
$\alpha_{-q}^{(n)}$ are determined by inital value conditions,
i.e. by $\Psi(0)$ and $\Psi^{\prime}(0)$. In the following we
choose the $u_{q}(x)$ to be normalized
\begin{equation}
\int_{0}^{d}dx\,u_{q}^{\ast}(x)u_{q}(x)=d/(2\pi).\label{uqnorm}%
\end{equation}

\subsection{The velocity operator has real eigenvalues at transmission resonances}

Any physical state is represented by a wave function which is an
element of the Hilbert space of the square integrable functions
$\mathcal{H}=L^2(\mathbb{R}^3)$. The Hermiticity of the velocity
(and the momentum) operator comes from the fact that the
normalization $\left\langle \Psi|\Psi\right\rangle=1$ leads to
$\Psi(\mathbf{x})=0$ as $|\mathbf{x}|\rightarrow\infty.$

The states we consider, Eq.~(\ref{Psi}), are scattering states
which do not belong to $L^2(\mathbb{R}^3)$. Therefore, we can not
expect that the velocity operator applied to any finite subdomain
is Hermitian. Nevertheless, we show that the expectation values of
the velocity operator inside the periodic structure are real
for resonant tunneling states.

The calculation is restricted to the one-dimensional region
between $x=0\,\,$and $x=L$ inside the FPS. The expectation value
of the velocity of any state $|\Psi\rangle$ is given by
$\left\langle \hat{v}\right\rangle =\left\langle \Psi|\hat{v}|\Psi
\right\rangle /\left\langle \Psi|\Psi\right\rangle$. In our case
the numerator can be denoted as $\langle
\Psi_{\text{FPS}}^{(n)}|\hat{v}\hat{P}_{\text{FPS}}|\Psi_{\text{FPS}}%
^{(n)}\rangle,$ where $\hat{P}_{\text{FPS}}$ is the projection operator onto
the space region of the FPS,
\begin{equation}
\hat{P}_{\text{FPS}}=\int_{0}^{L}dx\,|x\rangle\langle x|.
\end{equation}
Partial integration yields:%
\begin{align}
\langle\Psi_{\text{FPS}}^{(n)}|\hat{v}\hat{P}_{\text{FPS}}|\Psi_{\text{FPS}%
}^{(n)}\rangle &  =\left(  \frac{-ih}{m}\right)  \left.  |\Psi_{\text{FPS}%
}^{(n)}|^{2}\right|  _{0}^{L}+\nonumber\\
&  \langle\Psi_{\text{FPS}}^{(n)}|\hat{v}\hat{P}_{\text{FPS}}|\Psi
_{\text{FPS}}^{(n)}\rangle^{\ast}.\label{veloexpect-partint}
\end{align}
At resonances with $|t^{(n)}|=1$ the term $\left.  |\Psi_{\text{FPS}}^{(n)}%
|^{2}\right|  _{0}^{L}$ vanishes and we obtain:
\begin{equation}
\langle\Psi_{\text{FPS}}^{(n)}|\hat{v}\hat{P}_{\text{FPS}}|\Psi_{\text{FPS}%
}^{(n)}\rangle=\langle\Psi_{\text{FPS}}^{(n)}|\hat{v}\hat{P}_{\text{FPS}}%
|\Psi_{\text{FPS}}^{(n)}\rangle^{\ast},\label{vhermit}%
\end{equation}
i.e., the velocity has a real value.

\subsection{Expectation value of the velocity at resonance}

The expectation value of the velocity in the $n$-periodic system
at resonance is given by%
\begin{equation}
\langle\hat{v}_{\text{FPS}}^{(n)}\rangle=\frac{\langle\Psi_{\text{FPS}}%
^{(n)}|\hat{v}\hat{P}_{\text{FPS}}|\Psi_{\text{FPS}}^{(n)}\rangle}{\langle
\Psi_{\text{FPS}}^{(n)}|\hat{P}_{\text{FPS}}|\Psi_{\text{FPS}}^{(n)}\rangle
}.\label{veloexpectdef}%
\end{equation}
With the help of Eqs.~(\ref{PsiSLnbloch}) and (\ref{uqnorm}) the
numerator is given by%
\begin{equation}
\left\langle \Psi_{\text{FPS}}^{(n)}|\hat{v}\hat{P}_{\text{FPS}}%
|\Psi_{\text{FPS}}^{(n)}\right\rangle =\frac{L}{2\pi}\left(  \left|  \alpha_{q}%
^{(n)}\right|  ^{2}-\left|  \alpha_{-q}^{(n)}\right|  ^{2}\right)
v_{g}(q).\label{numerator of <v>}%
\end{equation}
The denominator of (\ref{veloexpectdef}) is given by%
\begin{equation}
\langle\Psi_{\text{FPS}}^{(n)}|\hat{P}_{\text{FPS}}|\Psi_{\text{FPS}}%
^{(n)}\rangle=\frac{L}{2\pi}\left(  \left| \alpha_{q}^{(n)}\right|
^{2}+\left|
\alpha_{-q}^{(n)}\right|  ^{2}\right)  .\label{denominator of <v>}%
\end{equation}
A detailed derivation of both equations is given in appendix
\ref{appendixint1} and \ref{appendixint2}, respectively. We stress that these
simple expressions hold only in the resonant case. From Eqs.
(\ref{veloexpectdef}) to (\ref{denominator of <v>}) we obtain the velocity
expectation value at resonance:
\begin{equation}
\langle\hat{v}_{\text{FPS}}^{(n)}\rangle=\frac{\left|  \alpha_{q}%
^{(n)}\right|  ^{2}-\left|  \alpha_{-q}^{(n)}\right|  ^{2}}{\left|  \alpha
_{q}^{(n)}\right|  ^{2}+\left|  \alpha_{-q}^{(n)}\right|  ^{2}}v_{g}%
=\frac{1-\left|  \alpha_{-q}^{(n)}/\alpha_{q}^{(n)}\right|  ^{2}}{1+\left|
\alpha_{-q}^{(n)}/\alpha_{q}^{(n)}\right|  ^{2}}v_{g}.\label{ratiovFPSvg}%
\end{equation}
An interpretation of this formula will be given in the next section.

\subsection{Determination of Bloch coefficients}

Next we have to determine the ratio $\left|  \alpha_{-q}^{(n)}/\alpha
_{q}^{(n)}\right|  $. For the calculation of the expectation value it would be
sufficient to consider only the wave function at resonance, nevertheless we
will derive a more general result which is valid for any $q$ value.

We rewrite the ansatz for the wave function (\ref{PsiSLnbloch}) in a reduced
form

\begin{subequations}
\label{PsiSLnbloch2WithAlpha}
\begin{flalign}
\Psi_{\text{FPS}}^{(n)}(x) &= \tilde{u}_{q}(x)\exp(iqx)+%
\tilde{\alpha}_{-q}^{(n)}\tilde{u}_{q}^{\ast}(x)\exp(-iqx),\label{PsiSLnbloch2} \\
\tilde{u}_{q}(x) &= \alpha_{q}^{(n)}u_{q}(x),\\
\tilde{\alpha}_{-q}^{(n)} &=\alpha_{-q}^{(n)}/\left(  \alpha_{q}^{(n)}\right)^{\ast}. \label{alphatilde}
\intertext{By formally replacing $q$ with $-q$ and taking the complex conjugate in Eq.~(\ref{alphatilde})} 
\tilde{\alpha}_{-q}^{(n)} &= 1/\left( \tilde{\alpha}_{q}^{(n)} \right)^{\ast} \label{alphatildeid}
\end{flalign}
\end{subequations}
is obtained.

As stated above, the remaining coefficient
$\tilde{\alpha}_{-q}^{(n)}$ is determined by the initial values
$\Psi(0)$ and $\Psi^{\prime}(0)$, obtained from the continuity of
the wave function and the probability current density across the
boundary at $x=0.$ To simplify the algebra, we make use of the
fact that these continuity conditions are inherently incorporated
in the transfer matrix approach. Therefore, to determine
$\tilde{\alpha}_{-q}^{(n)}$ we use the continuity of the
wavefunction at $x=0$ and two different representations of
$\Psi_{\text{FPS}}^{(n)}$ at $x=d.$ This way the derivative of the
wave function is not needed in an explicit form.
\cite{teq1comment} The continuity of the wavefunction (\ref{Psi})
at $x=0$ leads to
\begin{equation}
1+r^{(n)}=\tilde{u}_{q}(0)+\tilde{\alpha}_{-q}^{(n)}\tilde{u}_{q}^{\ast}(0).
\label{boundaryx0}%
\end{equation}
With Eq.~(\ref{PsiSLnbloch2}) the periodicity of the $\tilde{u}_{q}(x)$ gives the wavefunction at $x=d$:%

\begin{equation}
\Psi_{\text{FPS}}^{(n)}(d)=\tilde{u}_{q}(0)\exp(iqd)+\tilde\alpha_{q}%
^{(n)}\tilde{u}_{-q}^{\ast}(0)\exp(-iqd). \label{psixd1}%
\end{equation}
On the other hand $\Psi_{\text{FPS}}^{(n)}(d)$ can be obtained through the
transfer matrix (\ref{transfer matrix}) of the unit cell:%

\begin{align}
\Psi_{\text{FPS}}^{(n)}(d)  &  =(1 \quad 1)M^{-1}\left(
\begin{array}
[c]{c}%
1\\
r^{(n)}%
\end{array}
\right)  =\nonumber\\
&  =a^{\ast}-b^{\ast}+r^{(n)}(a-b). \label{psixd2}%
\end{align}
Again, we note that the continuity of the probability current density is fully
incorporated in this treatment in the reflection coefficient $r^{(n)}.$
Equations (\ref{boundaryx0}), and (\ref{psixd1}) together with (\ref{psixd2})
form two non-linear equations in $\tilde{u}_{q}(0)$ and $\tilde{\alpha}%
_{-q}^{(n)}$. The solution for $\tilde{\alpha}_{-q}^{(n)}$ is given by%

\begin{equation}
\tilde{\alpha}_{-q}^{(n)}=\frac{a^{\ast}-b^{\ast}-\xi+r^{(n)}(a-b-\xi)}%
{a-b-\xi+r^{(n)\ast}(a^{\ast}-b^{\ast}-\xi)},\label{alpha(n)1}%
\end{equation}
where $\xi=\exp(iqd)$ is introduced. From Eqs.~(\ref{transfermatrixrt}) to
(\ref{beta^{n}}) we obtain
\begin{equation}
r^{(n)}=\frac{b^{(n)\ast}}{a^{(n)}}=\frac{b^{\ast}U_{n-1}(\operatorname{Re}%
\{a\})}{aU_{n-1}(\operatorname{Re}\{a\})-U_{n-2}(\operatorname{Re}\{a\})}.
\end{equation}
Inserting into (\ref{alpha(n)1}) we get after some algebra%
\begin{equation}
\tilde{\alpha}_{-q}^{(n)}=\frac{a^{\ast}-b^{\ast}-\xi}{a-b-\xi}\,\frac
{a^{(n)\ast}}{a^{(n)}}=\frac{a^{\ast}-b^{\ast}-\xi}{a-b-\xi}\,\frac{t^{(n)}%
}{t^{(n)\ast}}.\label{alphankurz}%
\end{equation}

Consistent with (\ref{alphatildeid}), $\tilde{\alpha}_{-q}^{(n)}=%
1/\left( \tilde{\alpha}_{q}^{(n)}\right)^{\ast}$ is fulfilled.
In appendix \ref{simplewavef}, Eq.~(\ref{alphankurz}) is used to calculate
$\tilde{u}_{q}(x)$ and $\Psi_{\text{FPS}}^{(n)}(x).$ Further we prove in
appendix \ref{constantprobcurrent} the interesting identity%
\begin{equation}
\left\langle \Psi_{\text{FPS}}^{(n)}|\hat{v}\hat{P}_{\text{FPS}}%
|\Psi_{\text{FPS}}^{(n)}\right\rangle =j_{in}L, \label{vexp}
\end{equation}
where $j_{in}$ is the incident probability current. Eq.
(\ref{vexp}) is used for the calculation of $\alpha_{-q}^{(n)}$
and $\alpha_{q}^{(n)}$ in appendix \ref{alpha_q}.

As a consequence of (\ref{alphankurz}) the absolute
value of $\tilde{\alpha}_{-q}^{(n)}$ does not depend on $n$:
\begin{equation}
\left|  \tilde{\alpha}_{-q}^{(n)}\right|  = \left|%
 \frac{a^{\ast}-b^{\ast}-\xi}{a-b-\xi}\right|  .\label{absalpha}%
\end{equation}
This shows that the ratio of the amplitude of the left and right
going Bloch wave does not depend on the number of periods and is a property of the unit cell only.

Using $\operatorname{Re}\{a\}=\cos qd$, $|a|^{2}-|b|^{2}=1,$ and $\xi
=\exp(iqd),$ the absolute value squared of $\tilde{\alpha}_{-q}^{(n)}$ can be
simplified to:%
\begin{equation}
\left|  \tilde{\alpha}_{-q}^{(n)}\right|  ^{2}=\frac{\operatorname{Im}%
\{a\}+\sqrt{1-\operatorname{Re}^{2}\{a\}}}{\operatorname{Im}\{a\}-\sqrt
{1-\operatorname{Re}^{2}\{a\}}}.\label{absbeta^{2}}%
\end{equation}
This implies that at the edges of the allowed band $|\tilde{\alpha}_{-q}^{(n)}|$
reaches unity:
\begin{equation}
q=(0,\pi/d)+2p\pi/d \Leftrightarrow |\operatorname{Re}\{a\}|=1 \Leftrightarrow %
|\tilde{\alpha}_{-q}^{(n)}|=1.
\end{equation}

Taking $\left|  \tilde{\alpha}_{-q}^{(n)}\right|  =\left|  \alpha_{-q}%
^{(n)}/\alpha_{q}^{(n)}\right|  $ into consideration, inserting
(\ref{absbeta^{2}}) into (\ref{ratiovFPSvg}) gives finally
\begin{equation}
\langle\hat{v}_{\text{FPS}}^{(n)}\rangle=\frac{1-\left| \tilde{\alpha}%
_{-q}^{(n)}\right|  ^{2}}{1+\left|
\tilde{\alpha}_{-q}^{(n)}\right|  ^{2}}%
v_{g}=\frac{\sqrt{1-\operatorname{Re}^{2}\{a\}}}{-\operatorname{Im}\{a\}}%
v_{g}. \label{proof}%
\end{equation} However, this result is not true
for the off-resonant case when the first equality does not hold.
Nevertheless, the behaviour of
$\langle\hat{v}_{\text{FPS}}^{(n)}\rangle$ as a continuous function of
$E$ or $q$ is of interest, since any rational multiple of $\pi/d$ can be
obtained as $q_{\res}$, by choosing proper values for $j$ and $n$, see
Eq.~(\ref{qatresonance}).

For the sake of completeness, in the resonant 
case, where $q_{\res}$ is given by Eq.
(\ref{qatresonance}), and $t^{(n)}=\pm1$, Eq.~(\ref{alphankurz}) can be simplified to
\begin{equation}
\tilde{\alpha}_{-q_{\res}}^{(n)}=\frac{a^{\ast}-b^{\ast}-\xi}{a-b-\xi}.
\end{equation}

\section{Identity of tunneling time approach and velocity expectation value approach}
\label{sec:identity}
Comparing Eqs.~(\ref{ratiovresvg}) and
(\ref{proof}), we get the important identity
\begin{equation}
\langle\hat{v}_{\text{FPS}}^{(n)}\rangle=v_{\res}. \label{velocitiesareequal}%
\end{equation}
Consequently, the tunneling time approach and the velocity
expectation value approach are physically equivalent at resonance.
We use the term $v_{\res}$ as the electron velocity in resonant
tunneling through a FPS in the following. From the first identity
in Eq.~(\ref{proof}) follows, that the electron
velocity is bounded above by the group velocity $v_{g}:$%
\begin{equation}
v_{\res}\leq|v_{g}|. \label{v_le_vg}%
\end{equation}
Here the absolute value occurs, since the group velocity can also 
take negative values for positive $q$, e.g. for bands where the energy maximum
is at $q=0$, while the resonant velocity is always positive.

\subsection{Upper bound for the resonant tunneling velocity}

In fact, by squaring the outer equality in (\ref{proof}), an improved
inequality compared to (\ref{v_le_vg}) can be derived. 
We use that 
$|a|^2=|t|^{-2}\ge1$. 
Additionally, inside an allowed band $\operatorname{Re}^{2}\{a\}\le1$ holds. Therefore, 
$0\le1-\operatorname{Re}^{2}\{a\}\le\operatorname{Im}^{2}\{a\}$.
Now making use of the simple inequality 
$\frac{x}{y}\le \frac{x+z}{y+z}$ that holds for $0\le x\le y$ and $0\le z$
we obtain
\begin{equation}
\frac{1-\operatorname{Re}^{2}\{a\}}{\operatorname{Im}^{2}\{a\}}\leq\frac
{1}{\operatorname{Im}^{2}\{a\}+\operatorname{Re}^{2}\{a\}}=|t|^{2}.
\end{equation}
Finally, this proofs
\begin{equation}
v_{\res}\leq|t||v_{g}|. \label{vres_leq_tvg}
\end{equation}

In both formulas the equality holds for
$\operatorname{Re}\{a\}=0$, i.e. for $q=\pi/2d$, which is in the middle
of the band in $q$-space. This proves that the resonant tunneling velocity
can be much smaller than the group velocity, given a sufficient
small transmission amplitude of the unit cell. Furthermore,
towards the band edges where $\operatorname{Re}\{a\}$ approaches
$\pm1,$ the ratio $v_{\res}/v_{g}$ vanishes, see Eq.~(\ref{proof}).
Comparing with Eq.~(\ref{vgcaL/t}), taken from Ref. \onlinecite
{Landauer94}, we conclude that the matching of the wave functions
at the ends of the system can reduce the velocity considerably.

\subsection{Special case: Identity between $v_{\res}$ and
$|v_{g}|$}

From the outer equality in Eq.~(\ref{proof}) we can conclude that
the equality between $v_{\res}$ and $|v_{g}|$ holds if and only if
$|a|=|t|^{-1}=1$, i.e. at energies $E^{\prime}$ with
$|t(E^{\prime})|=1$ \emph{of the unit cell} of the periodic system:%
\begin{equation}
v_{\res}(E^{\prime}) =|v_{g}(E^{\prime})|\Leftrightarrow
|t(E^{\prime})|=1. \label{tvequalsgv}%
\end{equation}
A unit cell which is formed by a symmetric double barrier resonant
tunneling structure\cite{Ricco84,teq1comment2} possesses the
property $|t(E^{\prime})|=1$ at each transmission resonance energy
$E^{\prime}=E_{n}$. For this unit cell type the resonant tunneling
velocity $v_{\res}$ equals the magnitude of the group velocity
$v_{g}$ at all energies $E_{n}$. For energies different from the
$E_{n}$, $v_{\res}$ is smaller than $|v_{g}|$. The only unit cell
where $v_{\res}$ equals $v_{g}$ for all energies is the trivial
unit cell with $V(x)=0,$ for $0\leq x\leq d,$ since it has
$|t(E)|=1$ for all energies.

Further we can derive conditions from Eq.~(\ref{proof}) for the identity
between $v_{\res}$ and $|v_{g}|$, in terms of the coefficients $\alpha_{q}%
^{(n)},\alpha_{-q}^{(n)}:$%
\begin{align}
v_{\res}  &  =v_{g}\Leftrightarrow\alpha_{-q}^{(n)}=0,\\
v_{\res}  &  =-v_{g}\Leftrightarrow\alpha_{q}^{(n)}=0.
\end{align}
The resonant tunneling velocity equals the
magnitude of the group velocity if and only if the wave function has
only one Bloch component.

Considering Eq.~(\ref{tvequalsgv}) this can also be written as%
\begin{equation}
\alpha_{q}^{(n)}=0\vee\alpha_{-q}^{(n)}=0\Leftrightarrow|t|=1,
\end{equation}
yielding that the wave function inside a finite periodic potential
consists of only one Bloch wave if and only if the transmission of the
unit cell reaches unity.  Therefore, we obtain the picture, that the
Bloch wave moving into the left direction is built up by a coherent
superposition of the reflected parts of waves moving into the right direction.

This solves also the puzzle we mentioned in the introduction, namely,
that the phase delay time and the group velocity both originate from a
wave packet analysis, but lead to different results. Often the group
velocity is used when in good approximation no reflection occurs inside
the medium. In contrast, here, in the case that $|t|\neq1,$ there is
reflection in each period. These reflections are the origin of the
reduced velocity, compared to the group velocity.

\section{GaAs/AlGaAs superlattice} \label{sec:GaAs-SL} All presented results
are valid for arbitrary unit cells unless otherwise noted.  In this section the
results are applied to a periodic semiconductor heterostructure. Due to its
widespread use we choose a GaAs/AlGaAs superlattice (SL). Denoting the barriers
(AlGaAs) with B and the well regions (GaAs) with A, we can give a short
notation for the potential with $n$ periods: (BA)${}^n$. Assuming a stepwise
constant potential function in each layer the transfer matrix elements can be
calculated analytically. To keep the focus on the main topic, we will use the
following simplifications: (i) the effective mass $m^{\ast}$ is approximated to
be the same in GaAs and in AlGaAs, and (ii) the effective mass does not depend
on the energy. This simplifications can be avoided but this is left for future
work.  We present results for the lowest miniband. The analogous calculations
for higher minibands will be discussed elsewhere.

For a SL we get the following expressions for the matrix
elements\cite{Pacher03}

\begin{subequations}
\begin{align}
\operatorname{Re} \{a_{SL}\} &= \cosh \kappa L_{b}\cos kL_{w}-%
c_2\sinh\kappa L_{b}\sin kL_{w}, \label{ReaSL} \\
\operatorname{Im}\{a_{SL}\} &= -\cosh \kappa L_{b}\sin kL_{w}-%
c_2\sinh\kappa L_{b}\cos kL_{w}, \\
b_{SL} &= i c_1\sinh\kappa L_b \exp kL_w.   \label{bSL}
\end{align}
\end{subequations}

Here $L_b$ and $L_w$ are the thicknesses of the barrier and the well
layers, 
\begin{equation}
\kappa=\hbar^{-1}[2m^{\ast}(V_b-E)]^{1/2}, k =\hbar^{-1}(2m^{\ast}
E)^{1/2},\nonumber
\end{equation}
are the decaying electron wave vector in the barrier layer and the
electron wave vector in the well layer, respectively and
\begin{equation}
c_{1,2}=\frac{1}{2}(k\kappa^{-1}\pm \kappa k^{-1}).\nonumber
\end{equation}
The energy and the conduction band offset between GaAs and AlGaAs
are denoted $E$ and $V_b$, respectively.

The following SL parameters are choosen to be identical or nearly identical
to systems which were studied experimentally and theoretically in the
past\cite{Rauch98, Pacher01, Pacher03, Merc03}.
 
\begin{figure}[ptb]
\includegraphics[width=8.5cm]{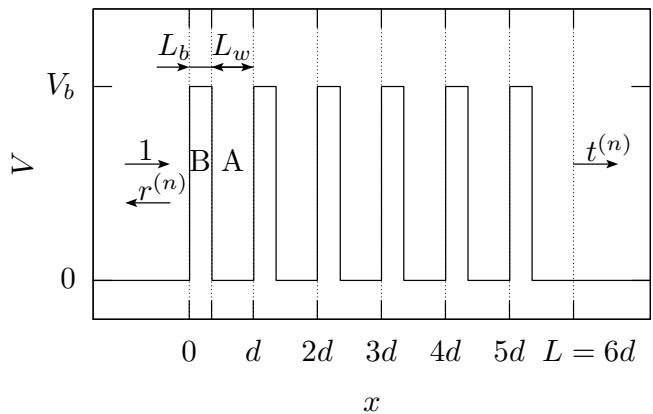}
\caption{Schematic potential of a superlattice of the form
(BA)$^{n}$. The number of periods $n$ is chosen to be 6.} 
\label{figpotential}%
\end{figure}

For the barrier and the well width we choose $L_b=2.5$~nm and $L_w=6.5$~nm,
resp. The barriers consist of Al${}_{.3}$Ga${}_{.7}$As, the wells of GaAs. 
The number of periods is $n=6$. A schematic of the resulting conduction
band is drawn in Fig.~\ref{figpotential}. The calculations are performed for
the $\Gamma-$valley of the conduction band for a temperature of $4$ K. As the
effective electron masses in both, GaAs and Al${}_{.3}$Ga${}_{.7}$As, we choose
$m^{\ast}=0.072m_0$, where $m_0$ is the free electron mass.  The conduction
band offset is $V_{b}=288$ meV. 

\begin{figure*}[ptb] 
\includegraphics[width=17cm]{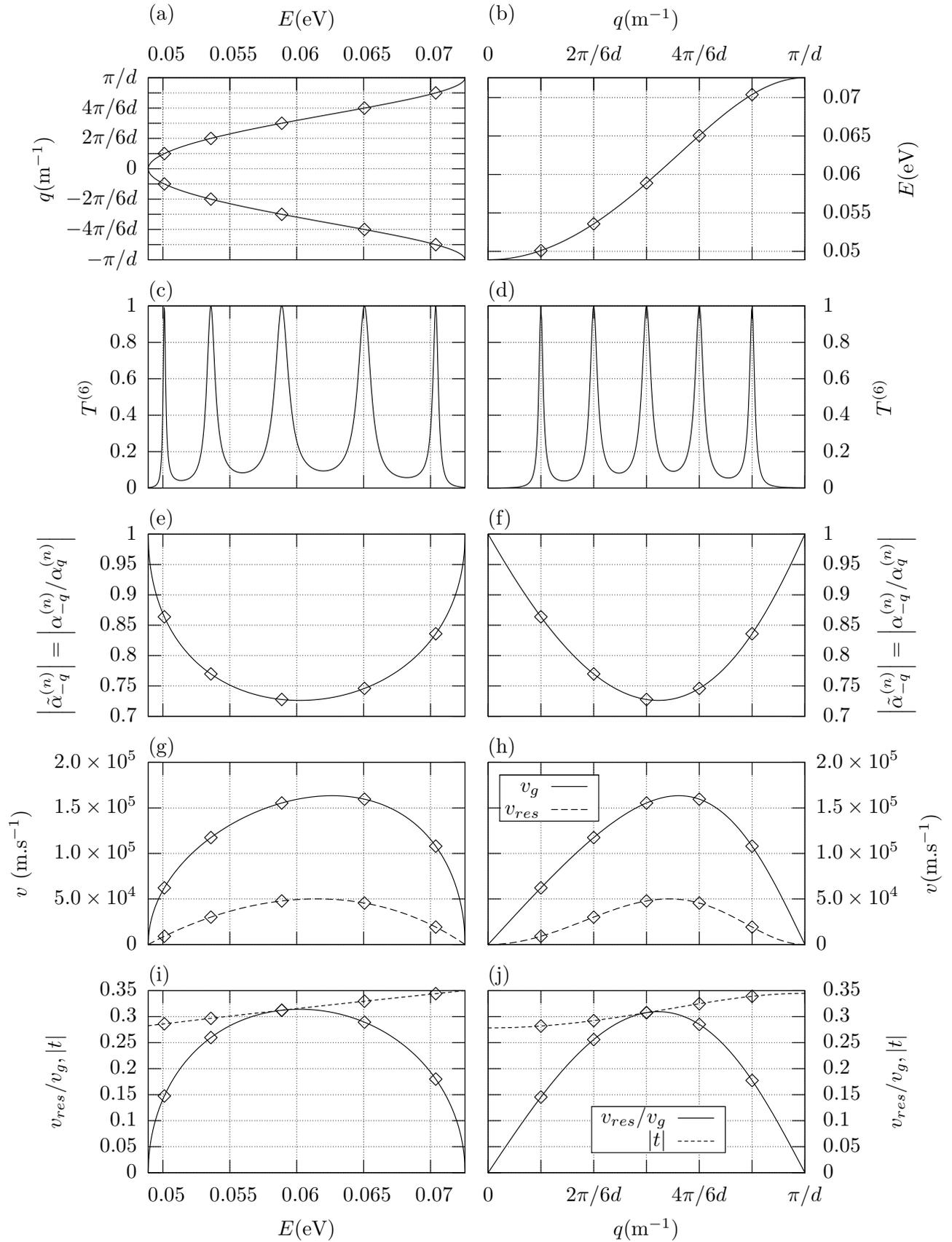} 
\caption{Shown
are relevant physical properties of superlattices. 
In the left column all functions are plotted vs. energy $E$, in the right column
vs. Bloch wave vector $q$.  Diamonds mark the positions of the resonant levels for
$n=6$ periods (band structure given in Fig.~\ref{figpotential}). By changing
the number of periods $n$, the lines connecting the diamonds are formed. The
dispersion relation, $q(E)$ (and $E(q)$), the transmission probability,
$T^{(6)}$, the normalized coefficient of the backwards propagating Bloch wave,
$|\tilde{\alpha}_{-q}^{(n)}|$, the group velocity, $v_g$, and the resonant
tunneling velocity $v_{\res}$, and the ratio of resonant tunneling and group
velocity of the periodic system together with the magnitude of the unit cell
transmission amplitude, $|t|$, are shown.
Details are given in the text.}
\label{figdisp_T_vg_vres} \end{figure*}

Because of the strong interdependencies between the dispersion relation,
transmission probability, group and resonant tunneling velocity, and the
ratio of the amplitudes of the $+q$ and $-q$ Bloch waves, all are shown
together in Fig.~\ref{figdisp_T_vg_vres}. All variables are plotted on
the left as a function of the energy $E$ and on the right as a function
of the Bloch wave vector $q$. 

For the first miniband the resonance energies are given by the smallest
solutions $E_j$ of Eqs.~(\ref{resonance finite SL}) and (\ref{ReaSL}),
i.e., by solving the transcendental equation \begin{equation}
\cos(j\pi/n) =\cosh\kappa L_{b}\cos kL_{w}-c_2\sinh\kappa L_{b}\sin
kL_{w}.  \end{equation} 

The diamonds mark the values at the resonance energies for the 
system with $n=6$ periods.
Changing the number of periods $n$  results in values that lie on the lines 
which connect the diamonds. In the limit $n\to \infty$ one would also obtain 
the continuous lines.

Figure~\ref{figdisp_T_vg_vres} (a) shows $q(E)$, and 
Fig.~\ref{figdisp_T_vg_vres} (b) shows its inverse, 
the dispersion relation $E(q)$. 
The dispersion relation is obtained from
\mbox{$\operatorname{Re}\{a_{SL}\}=\cos qd$}, where
$\operatorname{Re}\{a_{SL}\}$ is given by Eq.~(\ref{ReaSL}). 

In Fig.~\ref{figdisp_T_vg_vres} (c) and (d), the transmission 
probability, $T^{(6)}$, Eq.~(\ref{T^{(n)}}), is plotted.

From the plots of $E(q)$ and $T^{(6)}(q)$ the equidistant behaviour of
the resonant levels, Eq.~(\ref{qatresonance}), in the $q$-space can be
nicely seen. Of course, due to the nonlinear dispersion $E(q)$, this
behaviour is lost in energy-space.

Next, in Fig.~\ref{figdisp_T_vg_vres} (e) and (f), the magnitude of 
the amplitude ratio of the $-q$ and $+q$ Bloch waves, 
$|\tilde{\alpha}_{-q}^{(n)}|$, Eq.~(\ref{absalpha}), is shown. At the 
miniband edges this ratio reaches exactly unity, and it drops to a 
minimum around the middle of the miniband. 

Figure~\ref{figdisp_T_vg_vres} (g) and (h) show the group velocity,
$v_g$, Eq.~(\ref{vgroup}) (solid line), and the resonant tunneling
velocity, $v_{\res}$, Eq.~(\ref{resonancevelocity}) (dashed line). For
the $n=6$ case, the group velocity reaches its maximum in the fourth
resonance, while the resonant tunneling velocity reaches its maximum in
the third resonance. This shows again the fundamental difference between
both velocities.  It is interesting to note that the slope of
$v_{\res}(q)$ is zero at $q=0,\pi/d$, while the slope of $v_g$ has its
maximum value there. 

In Fig.~\ref{figdisp_T_vg_vres} (i) and (j), the ratio $v_{\res}/v_g$ and
the magnitude of the unit cell transmission amplitude, $|t|$, are
plotted. The velocity ratio is always smaller than
the single cell transmission, except for $q=\pi/2d$ when both are equal,
see Eq.~(\ref{vres_leq_tvg}). For the given parameters the velocity
ratio is below one third.

\section{Conclusions}

We calculated the resonant tunneling electron velocity in finite
periodic structures embedded in regions of constant potential in two
different ways and proved their identity.

The first method was based on the fact that $|t^{(n)}|=1$ leads to a
coincidence of all tunneling time definitions, turning up in the
literature. We used the phase time
$\tau_{\textrm{phase}}=\hbar\partial\arg t(E)/\partial E$ and the
natural definition $v=L/\tau$ to calculate the velocity in terms of the
unit cell transfer matrix elements and the unit cell transmission
amplitude, respectively, yielding Eqs.~(\ref{resonancevelocity}) and
(\ref{vres_t}).

Bloch's theorem was combined with the transfer matrix approach to
separate the wave function into two Bloch waves, which propagate in
opposite directions, Eq.~(\ref{PsiSLnbloch2}) together with either
Eq.~(\ref{absalpha}) or Eq.~(\ref{absbeta^{2}}). After proving that the
velocity operator is real at resonance, we calculated its expectation
value, Eq.~(\ref{proof}), as the second method.

Both results are completely identical, showing the physical equivalence
of the two approaches. The resonant tunneling velocity is in any case
smaller or equal to the group velocity. In addition, the resonant
tunneling velocity is smaller than or equal to the product of the group
velocity and the magnitude of the transmission amplitude of the unit
cell. Thus for unit cells with a small transmission amplitude both
velocities can differ considerably. We discussed that the Bloch wave
moving in the opposite direction of the incident electrons is due to
reflections inside all unit cells. These reflections are the
reason for the reduced velocity compared to the group velocity. At
energies, where the unit cell has a transmission of unity, only one
Bloch wave remains. Consequently the resonant tunneling velocity equals the
group velocity.

The group velocity is often used as the speed of the electrons inside a
superlattice. Our result points out that taking into account the
boundary conditions reduces the actual velocity considerably. 

In this context the often used relation "mean free path equals group velocity
times scattering time" seems questionable. We suggest that in mesoscopic
structures, where the mean free path exceeds the length of the periodic
structure it should be replaced by "\textit{mean free path equals resonant
tunneling velocity multiplied by scattering time}". The presented results
support the numerical findings of Ref. \onlinecite{Merc03}. 

We used electron waves throughout this paper. In analogy the main results hold
for the propagation of light waves, of phonons waves and so on in periodic
structures.

\begin{acknowledgments} Finally it is a pleasure to thank A. Wacker, J.
Burgd\"orfer and R. Dirl for interesting discussions and valuable
comments and to acknowledge financial support by the Austrian Science
Fonds (FWF) Grant No. Z24.
\end{acknowledgments}

\begin{appendix}

\section{Evaluation of $\langle\Psi_{\text{FPS}}^{(n)}|\hat{v}\hat
{P}_{\text{FPS}}|\Psi_{\text{FPS}}^{(n)}\rangle$\label{appendixint1}}

Expanding the integral with the help of Eqs.~(\ref{PsiSLnbloch}) and
(\ref{uqnorm}) first leads to \begin{widetext}
\begin{align}
\left\langle\Psi_{\FPS}^{(n)}|\hat{v}\hat{P}_{\FPS}|\Psi_{\FPS}^{(n)}%
\right\rangle= & \left|  \alpha_{q}^{(n)}\right|
^{2}\langle\Psi_{q}^{B}|\hat{v}\hat
{P}_{\FPS}|\Psi_{q}^{B}\rangle+\left|  \alpha_{-q}^{(n)}\right|
^{2}
\langle\Psi_{-q}^{B}|\hat{v}\hat{P}_{\FPS}|\Psi_{-q}^{B}\rangle\\
&  -\frac{i\hbar}{m}\left(
\alpha_{q}^{(n)}\alpha_{-q}^{(n)\ast}\int_{0}
^{L}\dx\,u_{q}(x)e^{iqx}\frac{\partial}{\partial x}\left(
u_{q}(x)e^{iqx} \right)  +c.c.\right)  .\nonumber
\end{align}
\end{widetext}
In resonance the Bloch wave vector $q$ is given by (\ref{qatresonance}), i.e.,
$q_{j}^{\res}=j\pi/L.$ For these $q_{j}^{\res}$ the last integral vanishes:
\begin{equation}
\int_{0}^{L}dx\,u_{q}(x)e^{iqx}\frac{\partial}{\partial x}\left(
u_{q}(x)e^{iqx}\right)  =\left.  \frac{1}{2}u_{q}^{2}(x)e^{i2qx}\right|
_{0}^{L}=0,
\end{equation}
due to the periodicity of both, the $u_{q}(x)$ and the complex exponential
function. Using the fact\cite{Kittel87} that
\[
\langle\Psi_{\pm q}^{B}|\hat{v}|\Psi_{\pm q}%
^{B}\rangle/\langle\Psi_{\pm q}^{B}|\Psi_{\pm q}
^{B}\rangle=v_{g}(\pm q)=\pm v_{g}(q),
\]
the normalization (\ref{uqnorm}) leads finally to Eq.
(\ref{numerator of <v>}).

\section{Evaluation of $\langle\Psi_{\text{FPS}}^{(n)}|\hat{P}_{\text{FPS}%
}|\Psi_{\text{FPS}}^{(n)}\rangle$\label{appendixint2}}

Expanding the integral with the help of Eqs.~(\ref{PsiSLnbloch}) and
(\ref{uqnorm}) leads to \begin{widetext}
\begin{align}
\left\langle\Psi_{\FPS}^{(n)}|\hat{P}_{\FPS}|\Psi_{\FPS}^{(n)}\right\rangle&
=\left| \alpha_{q}^{(n)}\right|
^{2}\langle\Psi_{q}^{B}|\hat{P}_{\FPS}|\Psi_{q} ^{B}\rangle+\left|
\alpha_{-q}^{(n)}\right| ^{2}\langle\Psi_{-q}
^{B}|\hat{P}_{\FPS}|\Psi_{-q}^{B}\rangle\nonumber\\
& +\left(
\alpha_{q}^{(n)}\alpha_{-q}^{(n)\ast}\int_{0}^{L}\dx\,u_{q}
^{2}(x)e^{i2qx}+c.c.\right)  .
\end{align}
\end{widetext}
Due to the periodicity of $u_{q}(x)$ and considering that $q_{j}^{\res}%
=j\pi/nd$, the last integral vanishes:%

\begin{align}
\int_{0}^{nd}dx\,u_{q}^{2}(x) & e^{i2qx} =\left(  \sum_{l=0}^{n-1}%
\exp(i2qld)\right)  \int_{0}^{d}u_{q}^{2}(x)e^{i2qx}dx\nonumber\\
&  =\sum_{l=0}^{n-1}\left[  \exp(i2\pi j/n)\right]  ^{l}\int_{0}^{d}u_{q}%
^{2}(x)e^{i2qx}dx=0,
\end{align}
where the last identity is due to the vanishing of the geometric sum. Then
together with the normalization (\ref{uqnorm}), we end up with Eq.
(\ref{denominator of <v>}).

\section{Simplified calculation of the wavefunction in the FPS}

\label{simplewavef} As a supplementary result, once
$\tilde{\alpha}_{-q}^{(n)}$ is calculated from (\ref{alphankurz}),
the values of $\Psi_{\text{FPS}}^{(n)}(x)$ \emph{in one period},
e.g. obtained with the transfer matrix approach, allow to
calculate the periodic function $\tilde{u}_{q}(x).$ Solving
(\ref{PsiSLnbloch2}) for $\tilde{u}_{q}(x)$ we obtain:
\begin{equation}
\tilde{u}_{q}(x)=\frac{\Psi_{\text{FPS}}^{(n)}(x)-\tilde{\alpha}_{-q}%
^{(n)}\left(  \Psi_{\text{FPS}}^{(n)}\right)  ^{\ast}(x)}{1-|\tilde{\alpha
}_{-q}^{(n)}|^{2}}\exp(-iqx).\label{uqdefined}%
\end{equation}
Using (\ref{PsiSLnbloch2}) and (\ref{alphankurz}), the
wavefunction in the \emph{entire FPS} can be obtained easily.

\section{Derivation of an identity and calculation of $\alpha_{q}^{(n)}$ and $\alpha_{-q}^{(n)}$}
\subsection{Identity - first method}
\label{constantprobcurrent} From Eqs.~(\ref{velocitiesareequal}),
(\ref{veloexpectdef}) and (\ref{vres=Lovertau}), i.e.%

\begin{equation}
\frac{L}{\tau_{\res}^{(n)}}=v_{\res}^{(n)}=\frac{\langle\Psi_{\text{FPS}}%
^{(n)}|\hat{v}\hat{P}_{\text{FPS}}|\Psi_{\text{FPS}}^{(n)}\rangle}{\langle
\Psi_{\text{FPS}}^{(n)}|\hat{P}_{\text{FPS}}|\Psi_{\text{FPS}}^{(n)}\rangle},
\end{equation}
and Eqs.~(\ref{dwelltimedef}) and (\ref{equalityofalltaus}), i.e.%
\begin{equation}
\tau_{\res}^{(n)}=\frac{1}{j_{in}}\langle\Psi_{\text{FPS}}^{(n)}|\hat
{P}_{\text{FPS}}|\Psi_{\text{FPS}}^{(n)}\rangle,
\end{equation}
we derive the interesting identity, valid at resonance,%
\begin{equation}
L=\frac{1}{j_{in}}\langle\Psi_{\text{FPS}}^{(n)}|\hat{v}\hat{P}_{\text{FPS}%
}|\Psi_{\text{FPS}}^{(n)}\rangle,\label{vL=<psivpsi>}%
\end{equation}
where the incoming probability current is given by $j_{in}=\hbar k/m.$ 

\subsection{Identity - second method}
This equation can also be derived directly from the conservation of electrical
charge in the form%
\begin{equation}
\frac{\partial}{\partial t}|\Psi(x,t)|^{2}+\frac{\partial}{\partial
x}j(x,t)=0,\label{continuityequation}%
\end{equation}
where $j$ is the one-dimensional probability current given by
\begin{equation}
j(x,t)=\operatorname{Re}\left[  \Psi^{\ast}\left(  -\frac{i\hbar}{m}%
\frac{\partial}{\partial x}\right)  \Psi\right]  .\label{probdensity}%
\end{equation}
Since in our case $|\Psi|^{2}$ and $j$ do not depend on time, Eq.
(\ref{continuityequation}) simplifies to
\begin{equation}
j(x,t)=\mathrm{const.}\label{jconst}%
\end{equation}
Now we integrate the constant probability current density, Eq.
(\ref{probdensity}), first for the wave function to the left, i.e.
$\Psi_{L}(x)=\exp(ikx),$ and second for the wave function at
resonance inside the FPS, i.e. $\Psi_{\text{FPS}}^{(n)}(x),$ both
over the intervall $[0,L].$ Then Eqs.~(\ref{jconst}) and
(\ref{vhermit}), yield also identity (\ref{vL=<psivpsi>}).

\subsection{Calculation of $\alpha_{q}^{(n)}$ and $\alpha_{-q}^{(n)}$}
\label{alpha_q}
Making use of Eq.~(\ref{numerator of <v>}), (\ref{vL=<psivpsi>}) and
(\ref{PsiSLnbloch2}) we can calculate $\alpha_{q}^{(n)}$ and $\alpha
_{-q}^{(n)}$. Choosing the (arbitrary) phase of $\alpha_{q}^{(n)}$, so that
$\alpha_{q}^{(n)}$ is real and positive, we obtain:%
\begin{align}
\alpha_{q}^{(n)}  &  =\left[  \frac{2\pi j_{in}}{\left(
1-|\tilde{\alpha}_{-q}^{(n)}|^{2}\right) v_{g} }\right]  ^{1/2},\\
\alpha_{-q}^{(n)}  &  =\tilde{\alpha}_{-q}^{(n)}\alpha_{q}^{(n)},
\end{align}
where $\tilde{\alpha}_{-q}^{(n)}$ is given by Eq.
(\ref{alphankurz}).

\end{appendix}
\bibliography{refs,comments}

\begin{thebibliography}{40}
\bibitem[{LMR()}]{LMRcomment}
\bibinfo{note}{To prevent some possible confusion we note that the transfer
  matrix can be also defined by $\left( \begin{array}[c]{c} A_{R}^{+}\\
  A_{R}^{-} \end{array} \right) =\hat M \left( \begin{array}[c]{c} A_{L}^{+}\\
  A_{L}^{-} \end{array} \right)$, for example used in Ref.
  \onlinecite{Pereyra02}. Of course $\hat M=M^{-1}$. The reader is asked to
  check carefully which definition is used when he consults the literature,
  since both occur frequently.}
\bibitem[{teq({\natexlab{a}})}]{teq1comment}
\bibinfo{note}{There remains one special case when $t=\pm1.$ For $t=\pm1,$ this
  simple approach does not work because the resulting equations are then linear
  dependent. In that special case one would have to use explicitely the
  continuitiy of the probability current density.}
\bibitem[{teq({\natexlab{b}})}]{teq1comment2}
\bibinfo{note}{We note that in this case $|t|=1,$ but $t\neq \pm 1$. Therefore
  it is no special case in the sense of footnote \onlinecite{teq1comment} and
  we can use the simple approach.}
\end{thebibliography}
\end{document}